\documentclass[11pt]{article}
\usepackage{amssymb,amsmath,cite}
\makeatletter
 
\long\def\@makefntext#1{\parindent 0cm\noindent
\hbox to 1em{\hss$^{\@thefnmark}$}#1}

\@addtoreset{equation}{section}
\def\appendix{\setcounter{section}{0}
        \def\thesection{Appendix.}
        \def\theequation{\Alph{section}.\arabic{equation}}}

\makeatother   

\begin{document}
\begin{flushright}
hep-th/0703115\\
March 2007\\
\end{flushright}
\begin{center}
{\Large\bf
 Transient Observers and Variable Constants\\ 
 {\Large\it or}\\[.6ex]
 Repelling the Invasion of the Boltzmann's Brains}\\
\vspace{.1in}
{S.~C{\sc arlip}\\
       {\small\it Department of Physics, University of California}\\
       {\small\it Davis, CA 95616, USA}\\
       {\small\it email: carlip@physics.ucdavis.edu}}
\end{center}
\begin{center}
{\large\bf Abstract}
\end{center}
\begin{center}
\begin{minipage}{5in}
{\small If the universe expands exponentially without end, ``ordinary
observers'' like ourselves may be vastly outnumbered by ``Boltzmann's
brains,'' transient observers who briefly flicker into existence as
a result of quantum or thermal fluctuations.  One might then wonder 
why we are so atypical.  I show that tiny changes in physics---for 
instance, extremely slow variations of fundamental constants---can 
drastically change this result, and argue that one should be wary of 
conclusions that rely on exact knowledge of the laws of physics in the 
very distant future.}
\end{minipage}
\end{center}

The fact that we observe an orderly universe is in part a characteristic
of the universe, but also in part a characteristic of us.  It is easy to
understand why we should see order---we are descended from a long line of
evolutionary ancestors who successfully navigated their local universe
long enough to reproduce, while competitors who were unable to correctly 
perceive the patterns of their environment were unlikely to have had descendants.
As Rees \cite{Rees}, Dyson et al.\ \cite{Dyson}, and Page \cite{Pagea,Pageb,%
Pagec} have pointed out, though, another kind of ``observer'' is possible: 
a ``Boltzmann's brain'' \cite{Albrecht}, a transient observer  appearing 
briefly as the result of a thermal or quantum fluctuation.\footnote{The term 
``Boltzmann's brain'' is a reference to Boltzmann's argument \cite{Boltzmann} 
that our ordered, low entropy universe could simply be a local thermal 
fluctuation in a much larger thermalized universe.  Given this possibility, 
it is easy to see that it is much \emph{more} likely for a fluctuation to 
merely produce an isolated ``observer.''}  The probability that a fluctuation 
in a given four-volume will produce anything we would call an observer is, 
of course, extraordinarily small.  But in an exponentially expanding, eternal 
universe, such ``Boltzmann's brains'' are inevitable, and under reasonable 
circumstances might vastly outnumber ordinary observers like ourselves.

There is no reason to expect that such transient observers would experience
an ordered universe, much less one with the particular order we see.  We might 
therefore ask why our observations are so atypical.  Whether this is a cause 
for worry is debatable---see, for instance, \cite{Hartle}---but at least for 
anthropic arguments, it seems to be a real concern: it is hard to argue that 
the universe should be suited to observers like us if typical observers are so 
completely different.

The simplest answer, of course, is that the observed accelerated expansion of 
our Universe may not be eternal.  While many quintessence 
models lead to an asymptotically constant dark energy density, for example, one 
can find others in which the effective cosmological constant eventually decays 
to zero; see, for instance, \cite{BBM,Cline,Kolda,Bento,Blais}.  Similarly, ``cyclic 
universe'' models \cite{Steinhardt} allow periods of exponential expansion that 
can end before the production of  ``Boltzmann's brains'' becomes significant.  
But observations are consistent with a true cosmological constant, and for models 
in which $\Lambda$ is constant---for example, those based on the string landscape 
\cite{Grana,Douglas}---the ``Boltzmann's brains'' problem may pose a serious 
challenge.   

A number of solutions to this puzzle have been proposed.  If the number of 
ordinary observers and the number of ``Boltzmann's brains'' are both infinite,
the relative probability depends on the choice of measure, and choices
exist for which ordinary observers predominate \cite{Vilenkin,Linde}.  Or 
perhaps the universe decays very rapidly \cite{Pagea,Pagec}---or, with a 
``holographic'' measure, not so rapidly \cite{Bousso}---and ``Boltzmann's 
brains'' have no time to appear.  Or perhaps a global conservation law
forbids the fluctuations that could lead to the appearance of observers 
\cite{Paged}.

One purpose of this note is to point out another simple possibility.  The
probability of a given thermal or quantum fluctuation depends on the value
of a number of dimensionless parameters such as the fine structure constant
and the electron-proton mass ratio.  If these parameters vary in time, even
at a rate \emph{much} slower than current experimental limits, the creation
of ``Boltzmann's brains'' may be strongly suppressed.  

Consider, for example, two of the ``Boltzmann's brains'' discussed by Page
in \cite{Pageb}.  A ``brief brain'' appears as a quantum fluctuation,
with energy $E$, characteristic length $r$, and action $S\sim Er/\hbar c$.  
For a ``brain'' containing $N$ nucleons, 
\begin{align}
E\sim Nm_pc^2 , \qquad
r\sim N^{1/3}a_0 = N^{1/3}\frac{\hbar^2}{m_ee^2}
\label{a1}
\end{align}
where $m_p$ is the proton mass, $m_e$ is the electron mass, and $a_0$ is the 
Bohr radius.  The action is thus
\begin{align}
S_{bb} \sim N^{4/3}\frac{m_p}{m_e\alpha} .
\label{a2}
\end{align}
(For $N$ on the order of $10^4$ times Avogadro's number, this agrees with
Page's estimate \cite{Pageb} of $S_{bb}\sim 10^{42}$.)
A ``long brain'' appears as a thermal fluctuation of de Sitter space, with
an action $S\sim\beta E$, where $\beta$ is the inverse de Sitter temperature;
that is,
\begin{align}
S_{lb} \sim 2\pi N \frac{m_pc^2}{\hbar H_\Lambda}
\label{a3}
\end{align}
where $H_\Lambda = c\sqrt{\Lambda/3}$ is the asymptotic Hubble constant.

The probability of a fluctuation goes roughly as $e^{-S}$, while the four-volume 
(at least for a given ``pocket universe'') grows as $e^{3H_\Lambda t}$.  The
number of ``Boltzmann's brains'' is thus 
\begin{align}
n_{bb}(t) &\sim \exp\left\{-N^{4/3}\frac{m_p}{m_e\alpha} + 3H_\Lambda t\right\} 
  \nonumber\\
n_{lb}(t) &\sim \exp\left\{-2\pi N \frac{m_pc^2}{\hbar H_\Lambda} + 3H_\Lambda t\right\}
. 
\label{a4}
\end{align}
These are, of course, very crude approximations, involving order-of-magnitude 
estimates in the exponents, but they give a reasonable qualitative picture.  In
particular, at present ($H_\Lambda t\sim1$) the numbers are tiny, but as $t$
increases, they grow without limit.

Implicit in this argument, however, is the assumption that the dimensionless 
parameters appearing in $S$---the fine structure constant, the electron-proton
mass ratio, etc.---are independent of time.  While this may be a reasonable starting
point, it is by no means a certainty.  In Kaluza-Klein theories, for example, such
constants depend on circumferences of compact dimensions, which need not be 
time-independent \cite{Forgacs,Weinberg}; in string theory, they depend on moduli 
whose dynamics can be quite complex \cite{Greene}.  Time-varying ``constants'' can 
appear in quintessence models \cite{Carroll,Wetterich}, in modifications of 
electromagnetism \cite{Bekenstein}, in variable speed of light models \cite{Magueijo}, 
and in brane world scenarios \cite{Brax}; for more references, see \cite{Uzan}.  
So it is worth examining the effect of relaxing the assumption of constancy.

Let us therefore suppose that our ``physical constants'' can vary in time.  Then
the exponent in the number $n_{bb}$ of ``brief brains'' given by (\ref{a4}) will 
not grow as long as 
\begin{align}
\frac{\dot\mu_1}{\mu_1} + \frac{\dot\alpha}{\alpha} < -3H_\Lambda N^{-4/3}\mu_1\alpha
  \sim -10^{-52}\,\mathit{yr}^{-1}
\label{a5}
\end{align}
where $\mu_1=m_e/m_p$.
This is about $36$ orders of magnitude below present experimental limits \cite{Webb,%
Srianand,Levshakov,Cingoz,Fortier,Dent,Ivanchuk}, and is not likely to be tested soon. 
For ``long brains'' the computation is slightly more delicate, since one should only 
consider time dependence of dimensionless constants \cite{Duff}.  Here, the condition 
that the universe be asymptotically anti-de Sitter should presumably be interpreted as
a statement of the constancy of $H_\Lambda$ in Planck units.  The relevant dimensionless
constant is then $\mu_2=m_p/M_{\hbox{\tiny\it Planck}}$, and the exponent in $n_{lb}$
will not grow in time as long as
\begin{align}
\frac{\dot\mu_2}{\mu_2} > \frac{3\hbar H_\Lambda^2}{2\pi Nm_pc^2} 
  \sim 10^{-80}\,\mathit{yr}^{-1}.
\label{a6}
\end{align}
I do not know of a clean experimental test of this quantity, but many of the searches
for variation of $G$---for instance, those based on stability of planetary orbits and
on stellar evolution---can be interpreted as limits on ${\dot\mu_2}/{\mu_2}$.  These
are reviewed in \cite{Uzan}; see also \cite{Sis}.  Big Bang nucleosynthesis also gives 
limits that depend on both $m_p$ and $ M_{\hbox{\tiny\it Planck}}$, which, barring 
unexpected cancellations, should limit variations of $\mu_2$ \cite{Uzan,Muller,Coc,Dentb}.  
The time dependence (\ref{a6}) is then, optimistically, at least $67$ orders of magnitude 
below present observational limits.

Let me next address a few technical points:\\[-4ex]
\begin{enumerate}\setlength{\itemsep}{-.5ex}
\item If $\alpha$, $\mu_1$, or $\mu_2$ vary in time, one might expect corresponding 
changes in $\Lambda$.  For instance, variations of $\alpha$ affect radiative
corrections, and, through those, particle masses and zero-point energies; a naive 
estimate leads to changes $\delta\Lambda/\Lambda$ much larger than $\delta\alpha/\alpha$
\cite{Donoghue}, large enough to rule out even the tiny change (\ref{a5}).  Changes 
in $\alpha$ or other constants also induce time dependence of the potentials for 
the corresponding moduli, which can again lead to variations in $\Lambda$ \cite{Banks} 
that can be eliminated only by fine-tuned cancellations.\footnote{The variations 
(\ref{a5}) and (\ref{a6}) actually fall comfortably below the ``no fine tuning'' 
limit of \cite{Banks}.}  Such arguments should serve as warnings against accepting 
``varying constants'' models too uncritically, but they are not decisive: we must
already require near-exact cancellation of the vacuum contribution to $\Lambda$,
and without knowing the mechanism of this cancellation we cannot say whether it
should be sensitive to variations in masses and couplings.  If the observed
near-zero value of $\Lambda$ is due to an accidental cancellation of vacuum energy
and a bare cosmological constant, for instance, then ``varying constants'' may be
sharply limited; if it is due to a dynamical mechanism, they may not be.

One may argue more directly for changes in $\Lambda$ if the evolution of masses and
coupling constants is directly determined by a quintessence field.  This certainly
need not be the case---string compactifications, for example, typically have large 
numbers of moduli with complex dynamics \cite{Grana,Douglas,Greene}, and any time 
dependence of vacuum energy may be quite different from that of the moduli that 
determine masses and couplings.  If $\Lambda$ \emph{does} vary and 
${\dot\Lambda}/\Lambda\sim{\dot\alpha}/\alpha$, this variation first becomes important 
in (\ref{a5}) at a time on the order of $10^{52}\,\mathit{yr}$.  Once this happens, 
the impact on the ``Boltzmann's brain'' problem depends on the sign of the variation: 
if $\Lambda$ increases, it may compete with the variation of $\alpha$, while if $\Lambda$ 
decreases, ``brain'' production is further suppressed.

\item In many models of varying constants, quantities such as $\alpha$, $\mu_1$, and
$\mu_2$ become asymptotically time-independent once the universe starts to accelerate
\cite{Sandvik,Barrowa,Barrowb}.  Such models will not solve the ``Boltzmann's brains''
problem, although if the relevant parameters do not become time-independent too quickly,
they may increase the amount of time available for another solution such as vacuum decay.
Note, however, that this asymptotic behavior depends on both the potential (it comes from 
the dominance of Hubble friction) and the relation between coupling constants and the 
underlying moduli.  As the Appendix shows, it is not hard to find models in which the 
asymptotic behavior is quite different.

\item I have so far been treating the time evolution of coupling constants and masses
as a purely classical phenomenon.  As we know from eternal inflation, quantum fluctuations
may be important \cite{Starobinsky}, and may disrupt the required monotonic evolution of
these parameters, pushing them back up the potential.  This effect depends on the shape 
of the potential and on the functional dependence of parameters such as $\alpha$ on the
moduli.  In particular, one might expect trouble for this paper's scenario if the
moduli approach equilibrium.  As shown in the Appendix, however, it is easy to construct 
a run-away potential for which the effect of quantum fluctuations is unimportant for 
the ``Boltzmann's brains'' problem.

\item Over long enough times, even slow variations of fundamental constants may lead 
to profound changes in physics.  If $1/\alpha$ grows linearly at the minimum rate 
allowed by (\ref{a5}), for instance, the electromagnetic interaction of two protons 
will become comparable to their gravitational interaction in about $10^{90}\,\mathit{yr}$.  
An increase in $\mu_2$ at the minimum rate allowed by (\ref{a6}) will have the same 
effect in about $10^{99}\,\mathit{yr}$.  It has been conjectured that gravity cannot 
become stronger than gauge interactions \cite{Arkani}; if this is the case, new 
physics would have to come into play.  In any case, it is unlikely in this scenario 
that ``Boltzmann's brains'' in the far future would look anything like the ``observers'' 
we now understand, and it is not clear that such objects would be possible at all.
\end{enumerate}

Should we thus conclude that our existence as observers implies a time dependence
of fundamental constants?  Presumably not.  I know of three broad solutions to the 
``Boltzmann's brains'' problem: dark energy density may not be (asymptotically) 
constant; our universe may tunnel to a new configuration quickly enough to avoid 
overproduction of transient observers; or masses, couplings, or other interactions 
may evolve in ways that reduce the ``brain'' production rate.  But within these 
categories, an enormous variety of particular solutions is available.  Dark energy 
could come from one of many inequivalent quintessence models.  Tunneling could take 
us to a universe with a different cosmological constant, as proposed by Page, but 
could also take us to a vacuum in which, say, transient observers fail to appear
because electroweak symmetry is unbroken.  Evolving couplings could suppress the
production of ``Boltzmann's brains'' as described here, or could lead to a universe 
in which, for instance, the existence of stable nucleons is no longer energetically 
favorable. 
 
Nor are these solutions mutually exclusive.  It is not hard to construct models, 
for example, in which couplings evolve too slowly to permanently suppress the production 
of ``Boltzmann's brains,'' but fast enough to greatly increase the time available 
for tunneling.  At the same time, we have only a limited understanding of the 
requirements for an ``observer'': the weakless universe \cite{Harnik} provides 
one illustration of how drastically ordinary physics could change without eliminating 
observers.

Rather, the lesson here is that we should be cautious about arguments that require
precise extrapolation of our present knowledge of physics to the very distant future.  
We have seen that truly tiny changes in fundamental constants, many orders of magnitude
below current observational limits, are enough to vitiate the ``Boltzmann's brains''
argument.  Given the uncertainties in our knowledge and the extreme sensitivity of 
the analysis to such uncertainties, it seems somewhat premature to draw conclusions 
about events $10^{42}$ years in the future.

\begin{flushleft}
\large\bf Acknowledgments
\end{flushleft}
I would like to thank Don Page and Andy Albrecht for stimulating discussions, and
Thomas Dent and John Barrow for helpful comments.
This work was supported in part by the U.S.\ Department of Energy under grant
DE-FG02-91ER40674.

\appendix
\section{Quantum fluctuations and varying constants}

As noted above, quantum fluctuations may potentially be significant in evaluating
the evolution of coupling constants.  Here, I will briefly explore this issue in 
a simple model.

Suppose the fine structure constant depends on a single scalar modulus 
$\varphi$.\footnote{While a varying $\alpha$ is itself a scalar field, it need not 
have an action with a standard, canonically normalized kinetic term.  Rather, it
will normally be a function of other canonically normalized scalar fields, with
a functional form that cannot be determined from general arguments, but must be
analyzed in particular models.}
The $\varphi$-dependent part of the action can be written as
\begin{align}
I = \dots + \int d^4x\sqrt{-g}\left[ \frac{1}{4}\alpha^{-1}[\varphi]F_{ab}F^{ab}
  + \frac{1}{2}g^{ab}\partial_a\varphi\partial_b\varphi - V[\varphi]\right] .
\label{b1}
\end{align}
Let us take the simplest dependence of $\alpha$ on the modulus, 
$\alpha^{-1}[\varphi] = \varphi/b$, so that (\ref{a2}) becomes
\begin{align}
S_{bb} \sim N^{4/3}\frac{m_p\varphi}{m_e b} = a\varphi.
\label{b2}
\end{align}
We shall assume---and later check---that the potential $V[\varphi]$ makes a negligible 
contribution to the cosmological constant, and that the evolution of $\varphi$ is
dominated by Hubble friction.  It is then known \cite{Starobinsky} that quantum 
fluctuations act as an effective white noise term in the classical equations of 
motion, and that the probability of finding a value $\varphi$ of the scalar field
is determined by the Smoluchowski equation
\begin{align}
\frac{\partial P(\varphi,t)}{\partial t} 
  = \frac{H^3}{8\pi^2}\frac{\partial^2 P(\varphi,t)}{\partial\varphi^2}
  + \frac{1}{3H}\frac{\partial\ }{\partial\varphi}%
  \left(\frac{dV}{d\varphi}P(\varphi,t)\right) .
\label{b3}
\end{align}

Let us now take $V[\varphi]$ to be a linear potential, $V[\varphi] = -k\varphi$, in
the range of $\varphi$ of interest.  Equation (\ref{b3}) is then equivalent to that
for Brownian motion in a constant gravitational field, and has the solution
\begin{align}
P(\varphi,t) = \sqrt{\frac{2\pi}{H^3t}}%
  \exp\left\{ -\frac{2\pi^2}{H^3t}(\varphi - {\bar\varphi}(t))^2\right\} ,
\label{b4}
\end{align}
where ${\bar\varphi}(t) = \varphi_0 + \frac{k}{3H}t$ is the classical solution.
The average number of ``Boltzmann's brains'' is easily computed.  From (\ref{b2}) 
and (\ref{b4}),
\begin{align}
\left\langle e^{-S_{bb}}\right\rangle = \int d\varphi P(\varphi,t)e^{-a\varphi}
  = e^{-{\bar S}_{bb}}e^{\frac{H^2a^2}{8\pi^2}Ht} ,
\label{b5}
\end{align}
where ${\bar S}_{bb}$ is the classical action for ``brief brains.''  The brain number
(\ref{a4}) thus becomes
\begin{align}
\langle n_{bb}\rangle(t)\sim\exp\left\{-N^{4/3}\frac{m_p}{m_e\alpha} + 3{\tilde H}t\right\}
 \qquad\hbox{with}\ {\tilde H} = H + \frac{H^2a^2}{24\pi^2} .
\label{b6}
\end{align}

The ``quantum correction'' $H^2a^2/24\pi^2$ goes as $1/b^2$, and is of order one for 
$b\sim 1\,\hbox{MeV}$.  We must still check the consistency of our assumption that 
$V[\varphi]$ does not contribute significantly to $\Lambda$.  To do this, note first 
that ${\dot V}\sim k^2/H$, and that $\Lambda\sim H^2/T_{\hbox{\tiny\it Planck}}^2$.
We thus want
\begin{align} 
\epsilon = \frac{1}{H}\frac{\dot V}{\Lambda} 
  \sim \frac{k^2T_{\hbox{\tiny\it Planck}}^2}{H^4} \ll 1 .
\label{b7}
\end{align}
But ${\dot\alpha}/\alpha = -{\dot\varphi}/\varphi \sim k/H\varphi$, so
\begin{align}
a = \frac{S_{bb}}{\varphi} \sim \frac{H}{k}S_{bb}\frac{\dot\alpha}{\alpha} 
  \sim \frac{H}{k}{\dot S}_{bb} \sim \frac{H^2}{k} ,
\label{b8}
\end{align}
where the final relation comes from our earlier condition that the
number of ``Boltzmann's brains'' not grow in time.  Thus
\begin{align}
\frac{H^2a^2}{24\pi^2} \sim \frac{H^6}{k^2} \sim 
  \frac{H^2T_{\hbox{\tiny\it Planck}}^2}{\epsilon} \sim \frac{10^{-120}}{\epsilon} .
\label{b9}
\end{align}
The potential $V$ can thus make a tiny contribution to the vacuum energy while
still allowing the quantum corrections to the classical evolution of $\alpha$
to be negligible.

This is, of course, a greatly oversimplified model, and I have not discussed
such issues as stability under radiative corrections.  Moreover, a linear
potential $V$ will eventually lead to a large change in the cosmological
constant, although perhaps not for $10^{130}\,{\mathit yr}$.  Note, though, that 
in one sense the model is rather conservative: I have made only minimal use of
the freedom to choose the function $\alpha^{-1}[\varphi]$.  Simply choosing 
$\alpha^{-1}\sim\varphi^2$ rather than $\alpha^{-1}\sim\varphi$ would be enough 
to guarantee that the growth in $S_{bb}$ eventually beats the Hubble expansion,
for instance, while more complicated forms would allow much greater freedom in
the choice of the potential $V$.


\small
\begin{thebibliography}{99}
\bibitem{Rees} M.\ J.\ Rees, {\it Before the Beginning: Our Universe and Others} 
  (Helix Books, 1997), p.\ 221.
\bibitem{Dyson} L.\ Dyson, M.\ Kleban, and L.\ Susskind, JHEP 0210 (2002) 011,
  hep-th/0208013.
\bibitem{Pagea} D.\ N.\ Page, hep-th/0610079.
\bibitem{Pageb} D.\ N.\ Page, JCAP 0701 (2007) 004, hep-th/0610199.
\bibitem{Pagec} D.\ N.\ Page, hep-th/0612137.
\bibitem{Albrecht} A.\ Albrecht and L.\ Sorbo, Phys.\ Rev.\ D70 (2004) 063528, 
  hep-th/0405270.
\bibitem{Boltzmann} L.\ Boltzmann, Nature 51 (1895) 413.
\bibitem{Hartle} J.\ B.\ Hartle and M.\ Srednicki, arXiv:0704.2630 [hep-th].
\bibitem{BBM} J.\ Barrow, R.\ Bean, and J.\ Magueijo, Mon.\ Not.\ Roy.\ Astron.\ 
  Soc.\ 316 (2000) L41, astro-ph/0004321.
\bibitem{Cline} J.\  M.\ Cline, JHEP 0108 (2001) 035, hep-ph/0105251.
\bibitem{Kolda} C.\ F.\ Kolda and W.\ Lahneman, hep-ph/0105300.
\bibitem{Bento} M.\ C.\ Bento, O.\ Bertolami, and N.\ C.\ Santos, Phys.\ Rev.\ D65 (2002) 
  067301, astro-ph/0106405.
\bibitem{Blais} D.\ Blais and D.\ Polarski, Phys.\ Rev.\ D70 (2004) 084008,
  astro-ph/0404043.
\bibitem{Steinhardt} P.\ J.\ Steinhardt and N.\ Turok, Phys.\ Rev.\ D65 (2002) 126003,
  hep-th/0111098.
\bibitem{Grana} M.\ Gra{\~n}a, Phys.\ Rept.\ 423 (2006) 91, hep-th/0509003.
\bibitem{Douglas} M.\ R.\ Douglas and S.\ Kachru, hep-th/0610102.
\bibitem{Vilenkin} A.\ Vilenkin, JHEP 0701 (2007) 092, hep-th/0611271.
\bibitem{Linde} A.\ Linde, JCAP 0701 (2007) 022, hep-th/0611043.
\bibitem{Bousso} R.\ Bousso and B.\ Freivogel, hep-th/0610132.
\bibitem{Paged} D.\ N.\ Page, J.\ Korean Phys.\ Soc.\ 49 (2006) 711,
  hep-th/0510003.
\bibitem{Forgacs} P.\ Forg{\'a}cs and Z.\ Horv{\'a}th, Gen.\ Rel.\ Grav.\ 11 (1979) 205.
\bibitem{Weinberg} S.\ Weinberg, Phys.\ Lett.\ B 125 (1983) 265.
\bibitem{Greene} B.\ Greene et al., hep-th/0702220.
\bibitem{Carroll} S.\ M.\ Carroll, Phys.\ Rev.\ Lett.\ 81 (1998) 3067, astro-ph/9806099.
\bibitem{Wetterich} C.\ Wetterich, JCAP 0310 (2003) 002, hep-ph/0203266.
\bibitem{Bekenstein} J.\ D.\ Bekenstein, Phys.\ Rev.\ D25 (1982) 1527.
\bibitem{Magueijo} J.\ Magueijo, Rept.\ Prog.\ Phys.\ 66 (2003) 2025, astro-ph/0305457.
\bibitem{Brax} P.\ Brax et al., Astrophys.\ Space Sci.\ 283 (2003) 627, hep-ph/0210057.
\bibitem{Uzan} J.-P.\ Uzan, Rev.\ Mod.\ Phys.\ 75 (2003) 403, hep-ph/0205340.
\bibitem{Webb} J.\ K.\ Webb et al., Phys.\ Rev.\ Lett.\ 82 (1999) 884, astro-ph/9803165.
\bibitem{Srianand} R.\ Srianand et al., Phys.\ Rev.\ Lett.\ 92 (2004) 121302,
  astro-ph/0402177.
\bibitem{Levshakov} S.\ A.\ Levshakov et al., astro-ph/0703042.
\bibitem{Cingoz} A.\ Cing{\"o}z et al., Phys.\ Rev.\ Lett.\ 98 (2007) 040801,
  physics/0609014.
\bibitem{Fortier} T.\ M.\ Fortier et al., Phys.\ Rev.\ Lett.\ 98 (2007) 070801.
\bibitem{Dent} T.\ Dent, JCAP 0701 (2007) 013, hep-ph/0608067.
\bibitem{Ivanchuk} A.\ Ivanchuk et al., Astron.\ Astrophys.\ 440 (2005) 45,
  astro-ph/0507174.
\bibitem{Duff} M.\ J.\ Duff, hep-th/0208093.
\bibitem{Sis} P.\ Sisterna and H.\ Vucetich, Phys.\ Rev.\ D 44 (1991) 3096.
\bibitem{Muller} C.\ M.\ M{\"u}ller, G.\ Sch{\"a}fer, and C.\ Wetterich, Phys.\ Rev.\ 
 D 70 (2004) 083504, astro-ph/0405373.
\bibitem{Coc} A.\ Coc et al., astro-ph/0610733.
\bibitem{Dentb} T.\ Dent and M.\ Fairbairn, Nucl.\ Phys.\ B653 (2003) 256, hep-ph/0112279.
\bibitem{Donoghue} J.\ F.\ Donoghue, JHEP 0303 (2003) 052, hep-ph/0101130.
\bibitem{Banks} T.\ Banks, M.\ Dine, and M.\ R.\ Douglas, Phys.\ Rev.\ Lett.\ 88 (2002)
  131301, hep-ph/0112059.
\bibitem{Sandvik} H.\ B.\ Sandvik, J.\ D.\ Barrow, and J.\ Magueijo, Phys.\ Rev.\ Lett.\
  88 (2002) 031302, astro-ph/0107512.
\bibitem{Barrowa} J.\ D.\ Barrow, J.\ Magueijo, and H.\ B.\ Sandvik, Phys.\ Lett.\ B541 
  (2002) 201, astro-ph/0204357.
\bibitem{Barrowb} J.\ D.\ Barrow and D.\ F.\ Mota, Class.\ Quant.\ Grav.\ 19 (2002) 6197,
  gr-qc/0207012.
\bibitem{Starobinsky} A.\ A.\ Starobinsky, in \emph{Field Theory, Quantum Gravity and 
  Strings}, edited by H.\ J.\ de Vega and N. S{\'a}nchez (Springer, 1986), Lect.\ Notes 
  Phys.\ 246 (1986) 107.
\bibitem{Arkani} N.\ Arkani-Hamed et al., hep-th/0601001. 
\bibitem{Harnik} R.\ Harnik, G.\ D.\ Kribs, and G.\ Perez, Phys.\ Rev.\ D74 (2006) 035006,
  hep-ph/0604027.
\end{thebibliography}
\end{document}